\documentclass[JIP]{ipsj}

\usepackage{amsmath}
\usepackage{graphicx}
\usepackage{latexsym}
\usepackage{cite}
\usepackage{url}

\usepackage{framed}

\usepackage{color}
\definecolor{darkblue}{rgb}{0.0,0,0.5} 

\usepackage{preprint}
\PROheadtitle{Submitted to Journal of Information Processing}

\def\Underline{\setbox0\hbox\bgroup\let\\\endUnderline}
\def\endUnderline{\vphantom{y}\egroup\smash{\underline{\box0}}\\}
\def\|{\verb|}

\received{2011}{7}{1}
\accepted{2011}{11}{5}

\usepackage[varg]{txfonts}
\makeatletter%
\input{ot1txtt.fd}
\makeatother%

\begin{document}

\title{Service Dependability with Continuously Revised Assurance Cases by Multiple Stakeholders: A Case Study}

\affiliate{YNU}{
Graduate School of Computer Engineering, 
Yokohama National University,
79-1 Tokiwadai, Hodogaya-ku, Yokohama 240--8501 JAPAN}

\author{Kimio Kuramitsu}{YNU}[kimio@ynu.ac.jp]

\begin{abstract}
Recently, assurance cases have received much attentions in the field of software-based computer systems and IT services. However, software very often changes and there are no strong regulations for software. These facts are main two challenges to be addressed in software assurance cases. We propose a development method of assurance cases by means of continuous revision  at every stage of the system lifecycle, including in-operation and service recovery in failure cases. The quality of dependability arguments are improved by multiple stakeholders who check with each other. This paper reported our experience of the proposed method in a case of the ASPEN education service. The case study demonstrate that the continuos updates create a significant amount of active risk communications between stakeholders. This gives us a promising perspective for the long-term improvement of service dependability with the lifecycle assurance cases. 
\end{abstract}


\maketitle


\section{Introduction}


{\em Assurance cases} are documentation-based engineering with structured arguments on the safety and dependability. Originally, assurance cases have been developed in the field of safety engineering for public transportation and industrial plant, and have been broadly adopted as a documentation standard to regulators\cite{Bloomfield2010}. The regulators especially in EU countries have made the assessment of assurance cases and then approved the developed system prior to its operation. 

Recently, due to the increased demands of the safety and dependability in software, many developers are interested in the application of assurance cases for software. However, software often changes over time, and even needs to change after the approval of the regulation. The emerging style of the DevOps\cite{OSD} development suggests that it would be difficult to separate development from the service operations. These natures make it difficult for a regulator to asses assurance cases, thereby resulting in the absence of strong regulators for software in general.

We propose a new development method of assurance cases for software-based IT services. The proposed method are based on two main ideas: lifecycle maintenance and stakeholder cross-reviewing. 

First, our aim for the use of assurance cases is to share dependability arguments between multiple stakeholders (e.g., developers, operators, and even users). We attempt to maintain assurance cases throughout a broader perspective of lifecycle including both development and operation. Naturally, we allow multiple stakeholders to revise assurance cases even in the post-development phases.

The absence of strong regulator is another challenge in terms of transferring high confidence with software assurance cases. To improve the confidence, we integrate a mechanism of several incentives to avoid faulty claims and evidence. 
The mechanism of incentives is based on the use of accountability concept with the rebuttals and cross-reviewing.

This paper reports our experimental experience on the development of the ASPEN online education system and its assurance cases. Multiple stakeholders, including developers, operators, and users, have participated into dependability arguments to avoid system failures. The arguments were written in GSN\cite{GSN}, a standard notation of assurance cases, and were shared to each others. During the experimental period of the ASPEN service, service failures unfortunately occurred although we made many extensive efforts with assurance cases, but the analysis of the failures gives us interesting insights.

In summary, findings throughout the experiment include:

\begin{itemize}
\item The initial cost for training stakeholders involved is not problematic, 
\item Assurance cases make all stakeholders explicitly convince the dependability, 
\item Dependability arguments grow even after the service is in operation, 
\item Reviewing by competing experts is strong and practical enough to check the faulty arguments
\end{itemize}

We conclude that assurance cases are a practical method in software-based IT services as to transfer dependability arguments from one stakeholder to the others through the lifecycle. We consider that such transferring is an important missing part for the long-term dependability of ever-changing software-based systems. 

The rest of the paper proceeds as follows. 
Section 2 is an introduction of assurance cases.  
Section 3 presents our basic ideas to develop assurance cases for software.
Section 4 describes  the ASPEN project. 
Section 5 examines the assurance cases that is developed in the ASPEN project.
Section 6 discusses lesson learnt. 
Section 7 reviews related work. 
Section 8 concludes the paper. 

\section{What is Assurance Cases?} 

\begin{figure}[tb]
\begin{center}
\includegraphics[width=85mm]{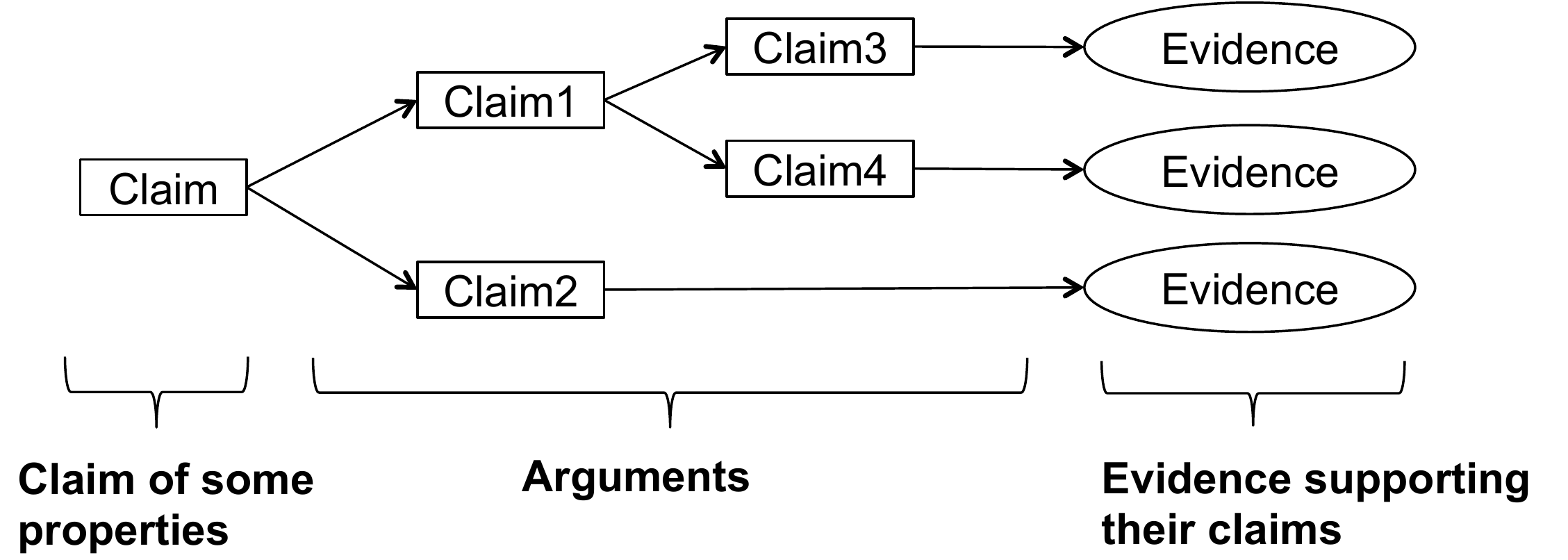}
\caption{Argument structure of assurance cases}
\label{fig:cae}
\end{center}
\end{figure}

Assurance cases are document-based engineering with structured arguments 
on the safety and the dependability. 
In this paper, we use the term dependability in a broader sense of the safety\cite{TDCS}. 
The documents are structured  for transferring the dependability confidence of products and services 
to others such as regulators and third-party organizations. 
To make the explicit the confidence, assurance cases are usually argued in a form of {\em Claim-Argument-Evidence}.
Figure \ref {fig:cae} illustrates the conceptual structure of assurance cases with the CAE arguments.

Let us suppose a claim, for example, that a system is adequately dependable to operate in
a given context. 
The argument explains the available evidence, showing how it reasonably supports the claim.
The top-most claim is decomposed into a series of sub-claims until these can be solved with evidence. 
Since the arguments make explicit the rational for the claim, they are more rigorous, justified, defensible, and transparent. 

Due to the high transparency of dependability arguments, 
assurance cases generally serves as an efficient {\em risk communication} tool between organizations. 
However, the most practically used scenario is transferring the developer's confidence to a regulator or a third-party company
to assess the conformance of dependability regulations\cite{IEEE12_Conformance}.
Through this assessment mechanism, the regulator forces the developer's product to meet their regulations, 
and then the users trust the developer's product due to the regulator's authority.

In contrast, the self-assessment of conformance or the absence of dependability regulations makes assurance cases  
{\em self-righteous} and their confidence weaken, resulting in an impractical case.

\section{Argumentation Architecture}

This section describes our ideas on how we use assurance cases software-based IT systems where there exists no strong regulator. 

\subsection{Sharing Dependability Arguments}

Our initial motivation comes from the risk of mis-communications between stakeholders such as developers and operators, who separately act in the distinct phases of the lifecycle. In other words, limitations discussed at the development time are quite useful for operators to deliver correct services, but they are unseen at the operation time. On the contrary, discussions at the operation time can be useful feedbacks for further development. Sharing such discussions focused on the dependability are extensively demanded to improve the long-term dependability of the products and services. 

Our aim for the use of assurance cases is sharing dependability arguments between stakeholders throughout the lifecycle. As introduced in Section 2, the arguments are well structured, and are more easy to convince the confidence due to the supporting evidence. This would suggest that assurance cases serve as a good foundation for sharing the focused knowledge, as well as risk communications. 

The argumentation architecture needs to slightly change when we attempt to apply it from one stakeholder-to-another to many-to-many stakeholders. 
First, the top claim must be a common goal and assumptions that are shared among all stakeholders. We decompose the common claim into sub-claims in a way that each stakeholder can separately lead the his or her acting parts of dependability arguments. 

The top claim is decomposed by {\em stages} in the lifecycle of products and services and then we decompose each stage claim by stakeholders if multiple acting stakeholders exist in the same stage. Each stakeholder has to provide available evidence that supports dependability claims that are part of the common goal.  

Staging in the lifecycle varies from project to project, but we refer to the following stages in this paper.

\begin{itemize}
\item Planing stage (requirement elicitation and architecting )
\item Development stage (coding and testing)
\item Operation stage (service design and failure recovery)
\item Evolution stage (change accommodation).
\end{itemize} 

Note that the stage decomposition above is based on the open system dependability \cite{OSD}, which we have proposed in JST/DEOS project. The uniqueness is the Evolution stage, where all stakeholders argue the continuously improvement of services beyond the lifetime of a single operated system.

\begin{figure}[tb]
\begin{center}
\includegraphics[width=85mm]{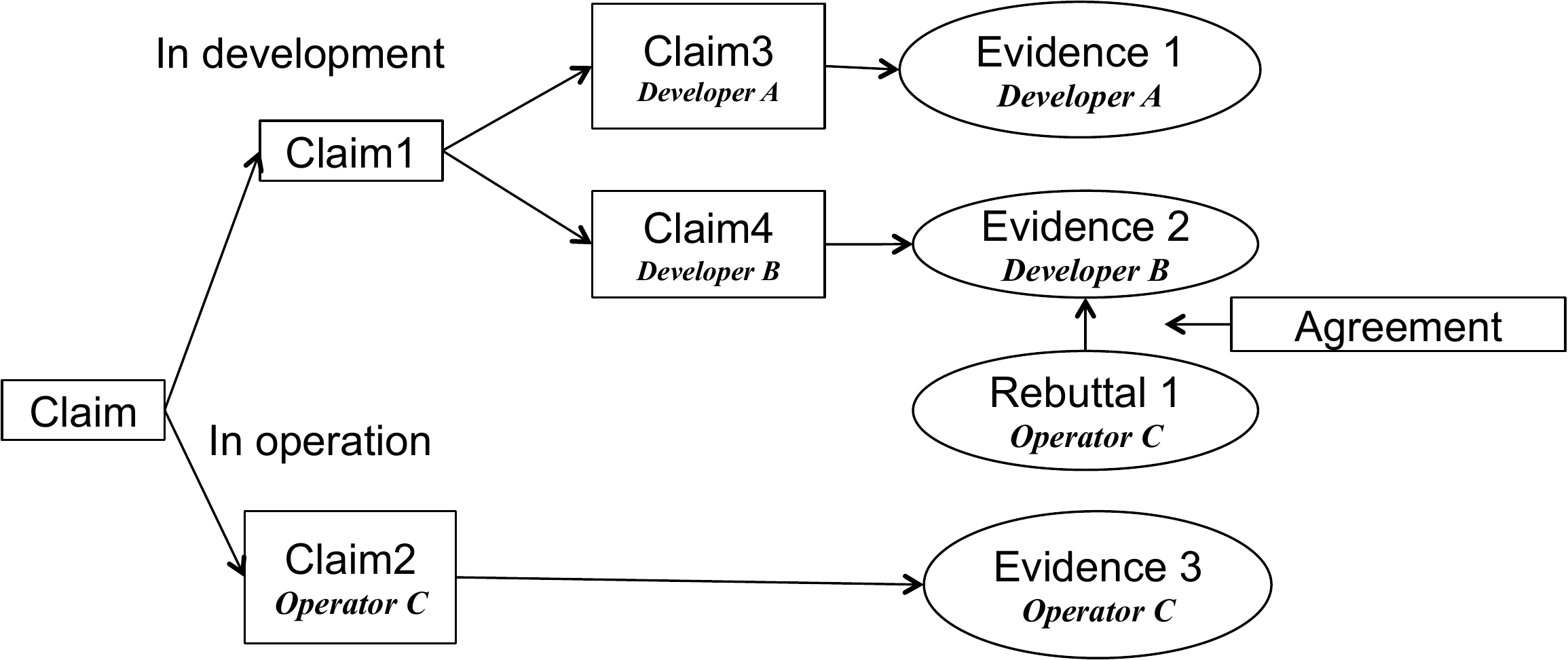}
\caption{Cross-Reviewing and Agreement} 
\label{fig:cae2}
\end{center}
\end{figure}

\subsection{Accountability, Rebuttals, and Revised}


Currently, most software-based IT services run under no regulation. As described in Section 2, the absence of strong 
regulators may reduce the practicality of assurance cases. This is a challenge to be avoided in practice.

The first idea is the use of the {\em accountability} concept\cite{PeerReview}; accountability is widely used to build trust and reputation among competing individuals and organizations by exposing failures. In contexts of assurance cases, we can integrate the accountability concept by recording the stakeholder identity at every element of assurance cases. That is, the stakeholder identity can be used to trace who claims or who gives faulty evidence when a problem occurs.
In general, this results in strong incentives to avoid faulty claim and evidence. 

In addition to the stakeholder accountability, we include a form of {\em rebuttal} into the dependability arguments. In contexts of assurance cases, the rebuttal means a challenge for the claim or an objection against the evidence, usually noted in the review process. In the assessment process, the rebuttals do not remain since they need to be solved prior to the certification. 
In the absence case of a regulator, the rebuttal is not so strong to enforce the modification. Unsolved rebuttals are regarded as conflicts. If the conflicts remain between stakeholders, the claim containing the rebuttal is in total regarded as the stakeholder agreements. Note that the rebuttals are recorded with the stakeholder identity for its accountability. 

Based on the recorded rebuttals, we use the cross-reviewing between stakeholders instead of the third-party reviewer, since stakeholders in part compete with each other (e.g., a developer wants reduced costs in the development but this is a potential risk to improper developed systems for others.) A faulty claim often becomes a potential risk for other stakeholders. 
Naturally, non-rebuttal claims are regarded as approved by all stakeholders with some sharable responsibility when a problem occurs. 
This also leads to another incentives to voluntarily make rebuttals for other's claim and evidence. Figure \ref{fig:cae2} illustrates our proposed argumentation architecture with lifecycle decomposition and stakeholder identities. 

More importantly, recall that our aim is to facilitate sharing dependability arguments between stakeholders, not to facilitate competition with each others. 
The developers and the operators can change software or service operations if they agree on given rebuttals. 
In addition, they are also allowed to revise the related assurance cases if their practice is changed.
This makes an iterative process, which enables us to better capture the ever-changing nature of software 
and would result in the dependability maintenance of changed software.

Note that we assume that all revised versions of assurance cases are referable by a proper version control system.

\section{ASPEN Project}

The ASPEN project is organized in the part of the JST/DEOS project in order to investigate the development method of assurance cases in reality. The ASPEN project includes not only assurance cases developed across different organizations, 
but also the development and the service operation of the ASPEN online education with our industrial partners. 
This section describes the experimental settings in the ASPEN project. 

\begin{figure}[tb]
\begin{center}
\includegraphics[width=75mm]{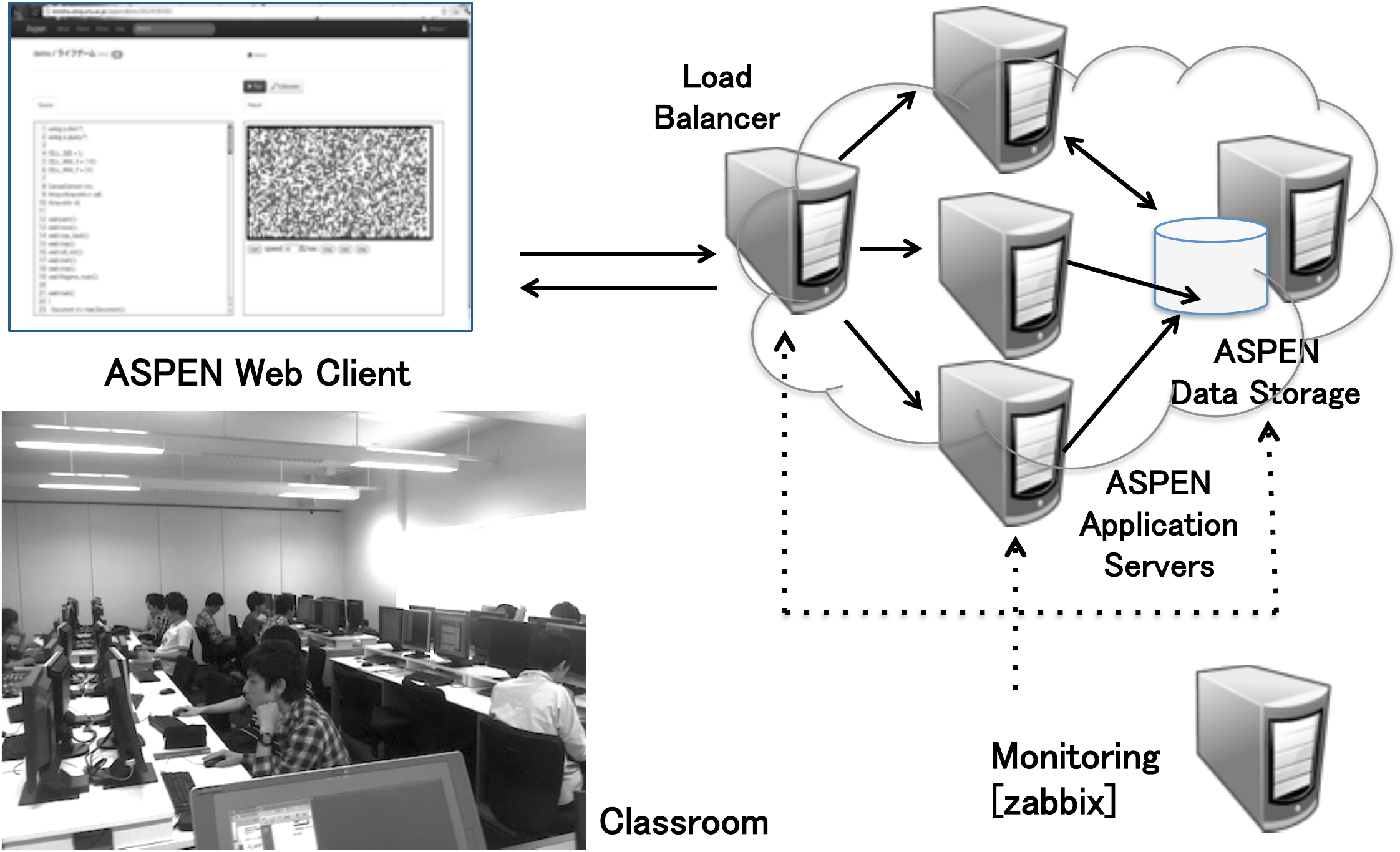}
\caption{ASPEN overview}
\label{fig:aspen}
\end{center}
\end{figure}

\subsection{System and Service Overview}

ASPEN is an online education system, which provides exercises on programming to students through the Web. Figure \ref{fig:aspen} is the overview of the ASPEN service. The students write code for assignments, and run it on the Web browser, and then submit it through a Web browser. The submitted assignments are stored on the ASPEN server, and subject to the university grading by an instructor. 

ASPEN is not a safety-critical system in such areas that assurance cases have been mainly developed so far. However, ASPEN involves several dependability attributes\cite{TDCS} that are commonly required in a wide class of 
software-based IT services.

\begin{itemize}
\item Availability -- The service will always be available to the users (i.e., students). 
\item Reliability	-- No hardware or software failures occur during provision of the service. 
\item Integrity -- Programming assignments supplied by the owner do not disappear. 
\item Privacy -- Personal information is not disclosed to unauthorized parties. 
\end{itemize}

\subsection{Stakeholders}
  

The earlier ASPEN was a small system that had been developed and operated for those studying in Yokohama National University. Several instructors had demanded to use the ASPEN in his or her class, and we planed to deliver 
the ASPEN service. At the same time, the dependability of the ASPEN system was a crucial issue for the instructors. 
To convince it we organized the co-development of assurance cases in parts of the JST/DEOS project.
We contracted the development and the service operation with two different firms. 

Here is a list of the stakeholders who were involved in the ASPEN project. In terms of the accountability, the stakeholder identity must be a personal name, but we identify them by the role name for readability.

\begin{itemize}
\item Owner -- the author
\item Developer -- a programmer working for an software development firm
\item Operator -- a system engineer with more that 10 year experiences on Web-based service operations
\item Users -- instructors and university students (i.e., those granted ASPEN access)
\end{itemize}
 
Note that the user that participated the dependability arguments is an instructor working for different institute from the authors'. 
 
\begin{figure}[tb]
\begin{center}
\includegraphics[width=70mm]{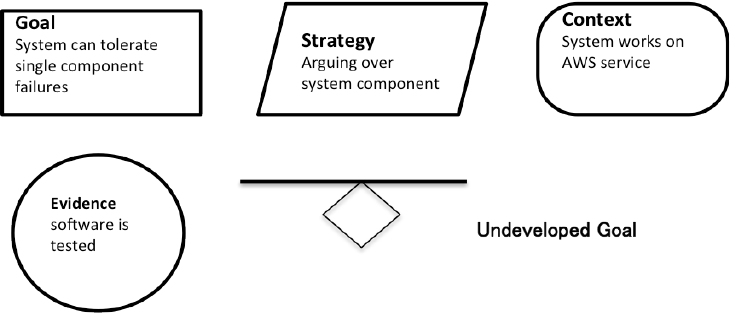}
\caption{GSN Elements}
\label{fig:gsn}
\end{center}
\end{figure}

\subsection{Goal Structuring Notation}

In the ASPEN project, we use {\em Goal Structuring Notation} \cite{GSN} to share assurance cases between stakeholders. The GSN notation is a standard argumentation notation of assurance cases in safety-critical industries. The GSN notation consists of four principle elements, goal (depicted in rectangle), strategy (parallelogram), evidence (oval), and context (rounded rectangle), as shown in Figure \ref{fig:gsn}. The goal element is used to state a claim that a system certainly has some desirable properties, such as safety and dependability. The evidence element is some piece of materials to support that the linked claim is true. The goal without linked evidence is called undeveloped. 

As in CAE, a goal is decomposed into sub-goals until a point is reached where claims can be supported by direct reference to available evidence. The strategy element is used to state a reason for claim decomposition. The context element is used to state an assumption (the system scope and the assumed properties). There is no special support for state the rebuttal. We regard the context element linked to the evidence as the rebuttal element.  

\begin{figure}[tb]
\begin{center}
\includegraphics[width=75mm]{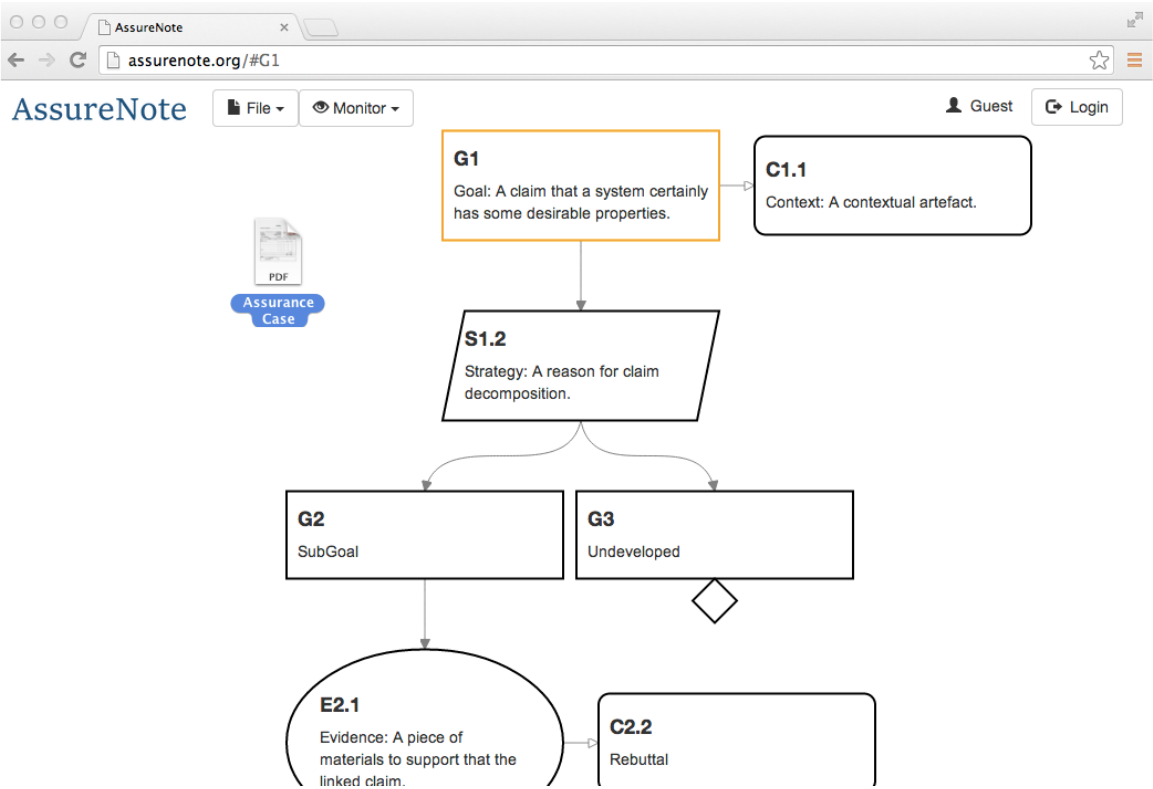}
\caption{AssureNote: a Web-based GSN authoring tool}
\label{fig:an}
\end{center}
\end{figure}

In the experiment, we use AssureNote\cite{AssureNote}, an Web-based authoring tool that allows multiple users to share the GSN document through the Web. Figure \ref{fig:an} is a screen shot of AssureNote. In AssureNote, all GSN elements are automatically recorded with the user identity under the version control of GSN elements. 
 
\subsection{Development Procedure}
 
In the experiment, we attempt the development of assurance cases in parallel with the development and service operation of the ASPEN system. All written assurance cases and other communications between stakeholders are in Japanese. 

In the Planing stage, the owner defines the top claim that includes dependability goals and assumptions, on which the ASPEN is assumed to deliver correct services for the users. Following the planning stage, we proceed the ASPEN with the following procedures: 

\begin{itemize} 
\item The developer claims that the developed system meets the owner's dependability goals with supporting evidence. 
\item The operator, the user, and the owner review the developer's claims and make agreements on the conflicted claims.
\item The operator, based on the developed system, claims that the operated system meet the owner's dependability goals with supporting evidence.
\item The developer, the user, and the owner review the operator's claims and make agreements on the conflicted claims.
\item Any stakeholder can revise the assurance cases if they find any insufficiency and defeats. 
\end{itemize}

As shown the above, we focus on the dependability related issues and avoid the general issues on software implementations and operation procedures. When we resolve the conflicted claims caused by rebuttals, we ask all stakeholders to meet together on the same place and then make agreements on the conflicts. 

\section{ASPEN Cases}

The ASPEN cases have been developed with a method that we have proposed in Section 3 and 4.
This section reports how the arguments are organized with a fragmentation of GSNs, excerpted from the developed assurance cases. 

\begin{figure}[tb]
\begin{center}
\includegraphics[width=75mm]{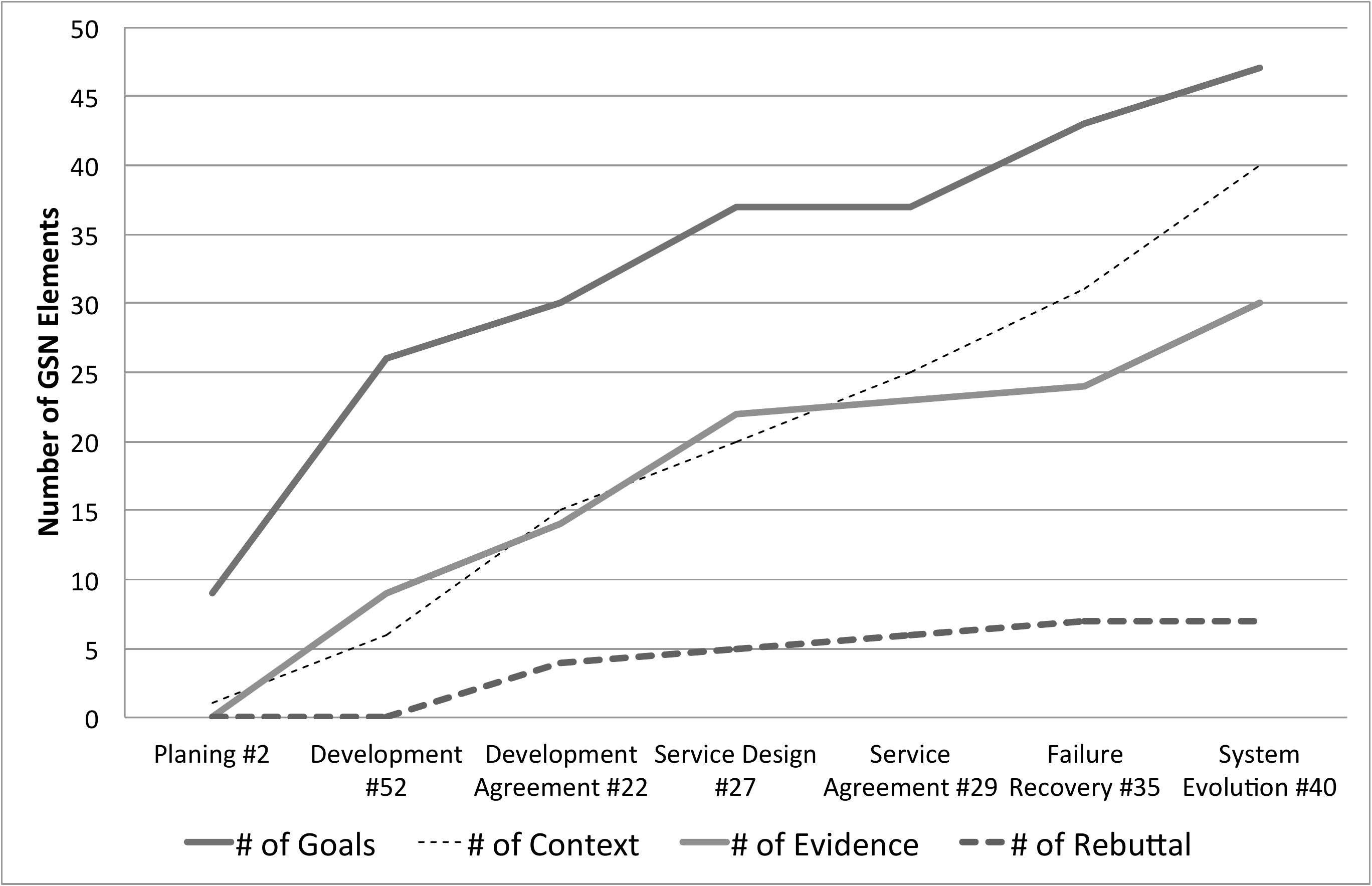}
\caption{Growth of assurance cases in GSN elements}
\label{fig:growth}
\end{center}
\end{figure}

\subsection{Overview}

First, we overview the statistics of the ASPEN cases. 
As we described in Section 4.3, the ASPEN cases are written in GSN. We first gave the top goal that was decomposed into the Development stage, the Service stage, and Evolution Stage with common assumptions. The GSNs were revised 40 times throughout the ASPEN project. 
Here we use \#$n$ to represent the $n$-th revision. 
The GSN document grew from 4 elements (at \#1) to 134 elements (at \#40). 
We identify the major revisions as follows:

\begin{itemize}
\item \#2 Initial Requirement
\item \#12 Development
\item \#22 Development Agreement 
\item \#27 Service design 
\item \#29 Service Agreement 
\item \#35 Failure recovery
\item \#40 Service evolution 
\end{itemize}

Figure \ref{fig:growth} shows the growth of the ASPEN cases in the number of GSN elements: goals, contexts, evidences, and rebuttals. The increase of contexts suggests the reduced ambiguity of the goals and the increase of evidence suggest the reduced uncertainty in dependability. 

In the remainder of this section we close up the development agreement (\#2-\#20), the service agreement (\#27-\#29), the failure recovery (\#35), and system evolution (\#40). 

\begin{figure}[tb]
\begin{center}
\includegraphics[width=75mm]{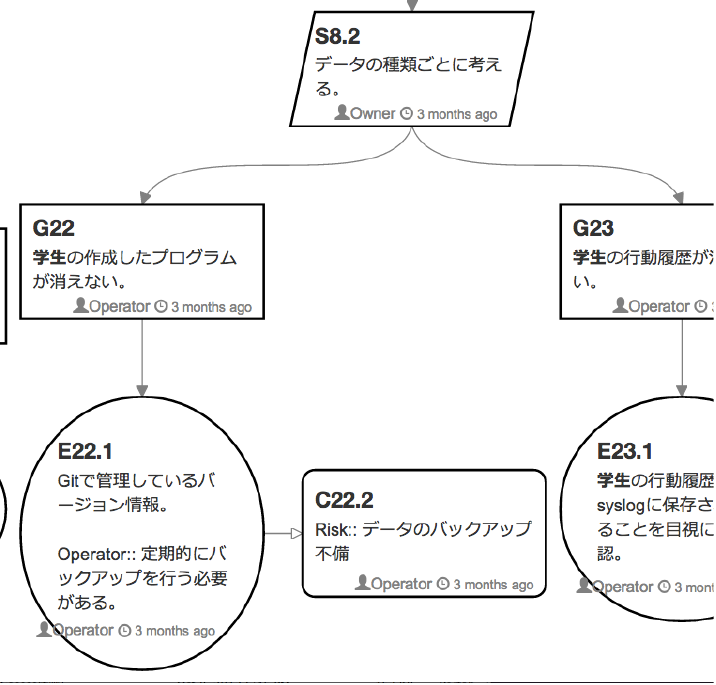}
\caption{Example of a rebuttal by the operator's review
({\sf G12}: The data never lose. Unillustrated in the figure.)
{\sf S8.2}: Arguing over type of data storage.
{\sf G22}: The submitted program data by the students never lose. 
{\sf E22.1}: The program data is stored the git repository. Operator agreement: periodical backups are necessary.
{\sf C22.2}: *Rebuttal* Lack of data backups.
{\sf C23}: The student activity data never lose.
{\sf E23.1}: The student activity is stored on the syslog storage. Tested by ... 
}
\label{fig:dev}
\end{center}
\end{figure}

\subsection{Development Agreement}

The developer started the argument with a claim "ASPEN is dependable", and decomposed it into sub-claims by dependability attributes (c.f., ASPEN is available, integral, safe, etc). 
The forms of evidence that the developer provided were mostly other external experience reports 
(collected from the Internet). Some goals remained the lack of evidence. Note that lack of evidence does not directly mean lack of dependability, but it does mean that uncertainty about dependability. 

Figure \ref{fig:dev} illustrates the fragment of developer's claim on that the software is integral in the data storage. 
One could consider that the evidence E22.1 was not reasonable evidence to support the claim G22. Likewise, the operator pointed out a risk of the hardware failure of disk storage. However there was no time enough to change the storage architecture. Instead, the operator agreed on the failure mitigation at the operation time.  Finally, the operator made 9 rebuttals for the developer's claim prior to the operations. 

Note that throughout the experiment we use the term "risk" instead of "rebuttal" to make it easier for participants to make  rebuttals. 

\subsection{Service Agreement}

In the operation stage, we focus only on the failure mitigation and failure recovery. The operator led the arguments as the domain stakeholder and used the failure avoidance pattern that consists of two parts:

\begin{itemize}
\item the symptom of a failure is detectable (by monitoring)
\item the detected symptom is mitigated before the service stopping
\end{itemize}

The completeness of failure avoidance analysis is important but not pursued in the experimentation in part because we want to measure the iterative effects of assurance cases during the operations.  

\begin{figure*}[tb]
\begin{center}
\includegraphics[width=145mm]{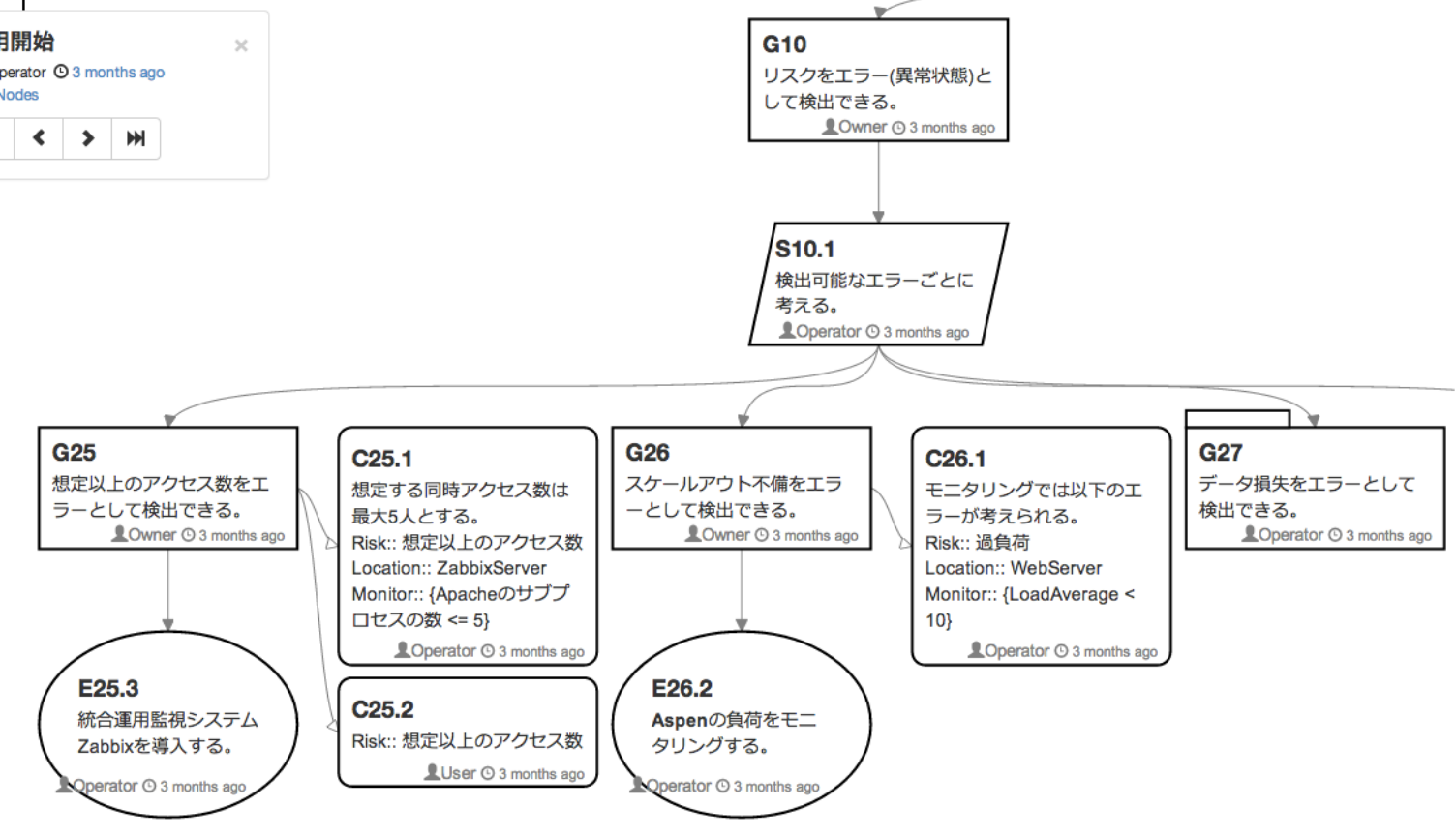}
\caption{Example of arguments on failure avoidance in operation:
{\sf G10}: Symptom of service failure is detectable as anomalies (measurable errors).
{\sf S8.2}: Arguing over type of errors.
{\sf G25}: Network traffic over the assumption is detectable.
{\sf C25.1}: The number of the assumed users is 5 at the same time. 
{\sf C25.2}: *Rebuttal* Exceeding the assumption
{\sf E25.3}:  Zabbix integrated monitoring tool is installed.
{\sf G26}: CPU overload is detectable.
{\sf C26.1}: CPU overload is defined by a threshold $w > 10$.
{\sf E26.2}: The CPU load average is monitored.
{\sf G27}: Data loss is detectable.
}
\label{fig:op}
\end{center}

\end{figure*}

Figure \ref{fig:op} is a fragment of assurance cases arguing the server's availability. Note that some parameters embedded are parameters for operation scripts\cite{DScript}. 
One could consider that the evidence is questionable but the user (and the owner) trusted the operator's claim with no doubt due to his expertise. They did not make any rebuttals against the operator, except for an explicit help-desk support in cases of service troubles. 

\subsection{Failure Recovery}

We encountered several service failures while we run the ASPEN service. 
Since unexpected failures means the insufficient arguments of failure mitigations, 
the operator revised the assurance cases based on stakeholder agreements.

Here we highlight a sever down failure, occurred in the middle of the classroom. 
This failure appeared only when students used ASPEN at the classroom. 
The system monitor indicated that the server operating system is unexpectedly down. 
At first, the operator suspected that there exists unknown bugs in the ASPEN system. 
Unfortunately, the developer found some bugs that might be related to the server down.
This is however a wrong way to recover the service. 
Three weeks later, the operator found an awful fact that 
the increase of traffic was unexpectedly fast, so that the servers could not scale out. 

In assurance cases,  the claim "the servers can scale out" was apparently self-rightours. 
The operator never tested in any actual traffic for evidence.
However, no body pointed out that this claim is insufficient.
After the problem occurred, the operator recognized that the scale-out approach was incapable of handling 40 classroom student access. The operator asked the instructor for all students 
not to access at the same time, 
although the instructor requested the operator to increase the servers as they had claimed on the assurance cases. 

Another serious problem has been found later in the context of top goal, 
which describes common assumptions on the ASPEN service. 
Originally, the ASPEN service is assumed for 100 students to submit their assignments from home computers. 
On this assumption, the maximum simultaneous access was estimated at most 5.
The number of students in the class room seems fewer than 100 students, but the density of access traffic differs from the estimation. That is, there is a bug at the top goal assumption.

The service was at last recovered by the increased server resources specially for the classroom use.  

\subsection{System Evolution}

The ASPEN project run two years. In the second year, we evolved the ASPEN system, based on the analysis of service failures and the users' demands in the first year. The resulting evolution includes:

\begin{itemize}
\item the adoption of the available open source solution such as Moodle to reduce in-house development
\item Transitioned computing environment to the Amazon Web Services (a standard cloud computing platform)
\end{itemize}

At the same time, we attempted to argue this system evolution based on the assurance cases. This is mainly because the ASPEN project was organized in parts of the JST/DEOS project, where the long-term dependability beyond the lifetime of each single system is a focused issue to be addressed. This attempt however let us recognize how difficult to organize the argument structure from scratch since the arguments on the development and the operation relies on the argument patterns. 

We could not draw two system evolutions as reasonable conclusions from the claim such that the system evolution is required.
On the other hand, we could claim that two system evolutions make more dependable in terms of availability and usability. 

\section{Lesson Learnt and Discussions}

This section describes lessons we learnt throughout the experiment and discuss the perspective of assurance cases in software-based IT services. 

\subsection{Development Costs}

{\em (Lesson 1). The initial training costs for introducing assurance cases is not problematic. }

For all the participants, the concept of assurance cases was new in such a way that they first suspected the benefits of assurance cases. We gave them about one-hour technical lecture on the GSN notation (not the concept of assurance cases itself). This short lecture is helpful enough to join dependability arguments in the project. No one had difficulty in reading the GSN documents. As a result, the training cost is not a primary issue for introducing the GSN-based arguments. 

{\em (Lesson 2). Argument patterns guide the well-structured arguments. }

Designing the structure of dependability arguments is acknowledged to be difficult in \cite{Bloomfield2010}. We gave prepared several patterns for dependability arguments, which come mainly from the academic resources in \cite{TDCS}. The participants were naturally guided to proven structure of arguments such as dependability attributes (e.g., reliability and integrity) and fault analysis. On the other hand, the system evolution could not be patterned before the development of assurance cases, since we did not have good reference to the system evolution in the literature. The resulting arguments seemed to be distractive in terms of the structures. The usefulness of argument patters is confirmed. 

{\em (Lesson 3). Providing evidence is the most costy part. }

The hardest part of developing assurance cases is a collection of supporting evidence. In part because the assurance cases are new for both the developers and the operators, they did not prepare any forms of evidence even if they claim them in the goals of the assurance case. For example, the developers preformed many functional tests as usual, but the tested results are far from those requested in the non-functional arguments. Both the developers and the operators don't want spend extra resources for evidence, the way of effective evidence transformation is still an open question. 

\subsection{Continuous Revision}

We examine the lifecycle revision of assurance cases. 
It is hard to evaluate how the dependability is improved from this single case study, but we can make a distinction from the traditional development of assurance cases. As shown in  Figure \ref{fig:growth}, the increased number of GSN elements suggests a fact that there are many lacked arguments at the development phase from viewpoints of the users who want service dependability. Continuous revision can follow up such lacked arguments even while the service is in operation. 
Note that a fact that our development is not intensively assessed suggests that we may drop many necessary arguments prior to the service operation. We need furthermore investigations. 


{\em (Lesson 4). Rebuttal is an enforced communication vehicle between stakeholders. }

Another finding is that rebuttals made at the design stage can be a source of the hazard analysis at the operation time. 
First, we consider that the operator's rebuttal makes the developer's responsibility explicit for the claimer if the related system failure happens. However, it was an unrealistic idea that the developer takes all responsibilities if there are annotated rebuttals on their evidence. Rather, some rebuttals are available at the failure mitigation of the next stage.
The rebuttals are used as such a serious communication vehicle between stakeholders.

\subsection{Cross-Reviewing and Its Quality}

We introduce the cross-reviewing between competing stakeholders an alternative solution to a regulator or a third-party assessment. We measure the reviewing activities by the number of revisions:

\begin{itemize}
\item 10 revisions for the development review 
\item 2 revisions for the operation review
\end{itemize}

The development review is more vigorous than the operation review. The difference comes from the reviewers' expertise. 

The developer and the operator apparently compete with each other in a such way that poorly developed systems can make it for the operator difficult to operate. In reality, the operator indicated several defeats of evidence given by the developer. The operator  never agreed with the lack of evidence, and often requested for too much evidence that almost exceeds the capability of the developer. In many cases, the owner needed to coordinate their conflicts. Without the coordinator we don't convince that they reached agreements. The quality of assurance cases are improved, as indicated by the growth of GSN elements. 

{\em (Lesson 5). Dependability guidelines, even not as strong as regulations, are demanded to make agreements in cases of conflicts. }
  
The operators and the users has the similar competing relationship, but the users couldn't suggest any weakness of service operations that is documented in assurance cases. This is mainly due to the lack of knowledge for system operations; the user trusted the operator's claim and evidence. However, this was not lucky for the operator since the user strongly requested the fulfillment of the operator's claims in cases of failures. Apparently, the strict review is better than the failure payoff. 

{\em (Lesson 6). Expert judgements are required in the review process. }

In conclusion, continuously revised assurance cases create active risk communications between stakeholders on the evidence basis. We consider that this approach leads to the long-term improvement of service dependability. 

\section{Related Work}

Our work builds on existing many excellent ideas for assurance cases, 
particularly based on 
assurance representations such as CAE\cite{Bloomfield2010} and GSN\cite{GSN,GSN2},
lifecycle developments \cite{DSN07_ABD,OSD}, 
argument patterns\cite{SoftwareSafetyArguments,SAFE11_Software}, 
and reviewing arguments\cite{GSNReview,IJCBS12_SoftwareSafety}.
The use of accountability (with stakeholder identity) is a little spice for the lifecycle maintenance of assurance cases.  


Recently there has been increasing interest in the use of assurance cases in software-based assurance. 
Boomfield and Bishop \cite{Bloomfield2010} discussed the current practice
 of safety and assurance cases for software-based systems.
Graydon et al. \cite{DSN07_ABD} proposed an approach
to integrating assurance into the development process by co-developing
the software system and its assurance case.
Hawkins et al extensibly study the assurance aspect of software process\cite{SoftwareProcess} and software evolution \cite{SAFE14_Software}.

Since the modern software is well modeled in design, 
there has been increasing interest in the model-driven approach to the development of software assurance case.
Denney and Pai shows an automated assembly from a lightweight formal models of software development\cite{SAFE12_CaseAssembly,AdvoCATE}. 
Hawkins et al\cite{HASE15_Weaving} shows how to generate an assurance argument for a system using
information extracted directly from design, analysis and development models of that system. 

Assurance information needs to be kept confidential in safety-critical systems\cite{IEEE13_OpenAssurance}. 
This leads to the very limited research materials for researchers. Recently, due to the increased research attention in assurance cases, many excellent practice reports have been published in the literature. The major reports include unmanned aircraft\cite{DSN12_SafetyCase}, automobile (ISO26262) \cite{DATE15_ISO26262}, autonomous vehicle and aircraft safety critical software\cite{SAFE11_Software}, generic infusion pump\cite{GIP}, pacemaker\cite{Pacemaker}, health IT service\cite{HealthIT}.  Our experience is a largely manual development and perhaps not well organized due to the lack of the base regulation but unique in terms of general software-based IT services.



\section{Conclusion}

Recently, assurance cases have received much attentions in the field of software-based computer systems and IT services. However, software very often changes and there are no strong regulations for software. These facts are considered to two challenges to be addressed in software assurance cases. We propose a development method by means of continuous revision for assurance cases at every stage of the system lifecycle, including in-operation and service recovery in failure cases. Instead of regulator, multiple stakeholders check dependability goals and their supporting evidence with each other. This paper reported our experience of the proposed method in a case of the ASPEN education service. The case study demonstrate that the continuos updates create a significant amount of active risk communications between stakeholders. This gives us the promising perspective for the long-term improvement of service dependability with the continuously updated assurance cases. At the same time, it is hard to draw a definitive conclusion from one studied case. In future work, we perform furthermore empirical studies on  software and IT services and establish dependability assurance guidelines for software.
   
\begin{acknowledgment}
This work is originally supported by the JST/CREST grant "Dependable Embedded Operating Systems for Practical Uses". The authors thank to all the DEOS research members, especially Mario Tokoro, Shuichiro Yamamoto, Yutaka Matsuno, Midori Sugaya, Yoshiki Kinoshita, Yoshiki Kinoshita, Makoto Takeyama, and Makoto Yashiro.

\end{acknowledgment}

\bibliographystyle{ipsjsort-e}
\bibliography{../bib/parser,../bib/mypaper,../bib/url,../bib/deos}  

\begin{biography}

\profile{Kimio Kuramitsu}{is an Associate Professor, leading the Software Assurance research group at Yokohama National University. His research interests range from programming language design, software engineering to data engineering, ubiquitous and dependable computing. He has received the Yamashita Memorial Research Award at IPSJ. His pedagogical achievements include Konoha and ASPEN, the first programming exercise environment for novice students. He earned his earned his B. E. at the University of Tokyo, and his Ph. D. at the University of Tokyo under the supervision of Prof. Ken Sakamura.}

\end{biography}

\end{document}